\documentclass{article}

\usepackage[final]{neurips_2022}

\usepackage[utf8]{inputenc} 
\usepackage[T1]{fontenc}    
\usepackage{hyperref}       
\usepackage{url}            
\usepackage{booktabs}       
\usepackage{nicefrac}       
\usepackage{microtype}      
\usepackage{xcolor}         

\newcommand\blfootnote[1]{%
  \begingroup
  \renewcommand\thefootnote{}\footnote{#1}%
  \addtocounter{footnote}{-1}%
  \endgroup
}

\title{Expansive Participatory AI: Supporting Dreaming within Inequitable Institutions}

\author{
  Michael Alan Chang$^\ast$\\
  UC Berkeley\\
  \texttt{michaelc@berkeley.edu} \\
   \And
   Shiran Dudy$^\ast$ \\
   CU Boulder\\
   \texttt{shiran.dudy@colorado.edu} \\
}

\begin{document}

\maketitle

\blfootnote{$^\ast$ Equal Contribution.}
\begin{abstract}
Participatory Artificial Intelligence (PAI) has recently gained interest by researchers as means to inform the design of technology through collective's lived experience. PAI has a greater promise than that of providing useful input to developers, it can contribute to the process of democratizing the design of technology, setting the focus on \textit{what} should be designed. However, in the process of PAI there existing institutional power dynamics that hinder the realization of expansive dreams and aspirations of the relevant stakeholders.   
In this work we propose co-design principals for AI that address institutional power dynamics focusing on Participatory AI with youth.
\end{abstract}

\section{Introduction}

As Artificial Intelligence (AI) has become immersed in our society, the value of Participatory AI (PAI) has increased, providing means to engage the public in the designing of AI products, systems, and spaces~\citep{van11participatory,harrington2020forgotten,arana2021citizen}.
PAI amplifies the voices of the collective and democratizes the process through which these products are being realized. While developing awareness around the value of PAI remains a crucial first step, past work has shown that many existing PAI processes are \textit{technocentric}, where the co-design process is driven by a need to develop a tool supported through an emergent capability of AI~\citep{martin2020participatory}. This priority of co-design is further reinforced by the powered nature of co-design where researchers and "researchees" have differential technological knowledge~\citep{pierre2021getting,scheall2021priority}. Taken together, existing co-design techniques broaden participation but largely fall short on empowering impacted communities in expansively imagining socially just futures~\citep{birhane2022power}.

In this paper, we argue that co-design processes that empower communities to imagine expansive futures with technology such as AI must support participants in first interrogating and desettling~\cite{Bang2012DesettlingEI} the often inequitable taken-for-granted \textit{ways of knowing} and \textit{ways of doing} that are baked into the institution that is the object of co-design (e.g., classroom, hospital, etc.) ~\citep{arnstein1969ladder,lugard1926dual,sloane2020participation}. Ultimately, co-designed technologies that operate within existing problematic institutional norms may address a near-term need for participants, while reifying deeper systemic issues that continually serve to oppress, dominate, and marginalize non-dominant communities~\citep{gray2019ghost,sloane2020participation}. 

Towards this goal, we first provide some general design principles that support creators of co-design spaces to desettle inequitable institutional practices. Next, we concretely illustrate these design principles in action through our experience co-designing educational AI-based tools with youth participants through our \textit{Learning Futures Workshop} which sought to work with high-school aged youth to imagine how AI has a role (if at all) in realizing their ideal collaborations within existing classroom spaces.

\section{Principals and Pillars for Expansive Co-Design with Youth}

\textbf{Desettling institutions through the Everyday and the Fantastical}: Creators of co-designer spaces should emphasize that institutional goals, norms, and practices, are socially constructed rather than "inherent". Researchers may explore this by working with youth to dream about how a common institutional practice may look if applied in a fantastical setting (e.g., allowing youth to explore how collaboration looks outside of the institution of schooling). \\
\textbf{AI as a means to an end}: Frame AI not as a means onto itself, but a means or a tool to help make possible an ideal, socially just institution. \\
\textbf{Removing concerns about AI Feasibility}: To de-normalize differential knowledge regarding what is technically feasible, remove AI feasibility as a constraint allowing youth to freely dream. Feasibility can often be addressed later in the process by achieving an \textit{approximation} of youth dreams through careful consideration. \\
\textbf{Support the complexity and contradictions of youth voices in co-design contexts}: Recognize that in the process of expansive dreaming, participants will often hold contradictory beliefs about what is needed (e.g., hopes for institutional transformation vs. needs within the existing institution). Facilitators must embrace the idea that co-design participants are \textit{learners}, gradually making sense of complex problems and solutions.

\section{Case Study: Educational Institution and the Learning Futures Workshop}

We designed a \textit{Learning Futures Workshop}, a workshop with the goal of surfacing youth's expansive hopes and dreams for ideal collaboration inside modern classrooms, and identifying the role of AI in making those dreams a reality. 

Youth, a key impacted community member in schools, daily operate within the modern of institution of schooling and it's associated instructional practices, goals, and hierarchies. These practices (e.g., grades, class periods, etc.) have been defined as the \textit{grammar of schooling}, which have proven to be resilient to reform efforts since their inception in the early 20th century~\cite{Tyack1994TheO}. While these classroom practices have been naturalized as simply "how things are", many have problematized the ideologies underlying the grammar of schooling. For instance, the grammar of schooling has been justified with the "logics of merit", which argue that students possess intrinsic characteristics that lead to success with the existing school system~\cite{Oakes2007RadicalCT}. In effect, the deficit framing of non-dominant students are baked into the institutional fabric of schooling. Past AI-based co-design in the education context has tended to reproduce ideologies present in the grammar of schooling~\cite{LFW}.

Consistent with the design principles in section 2, we de-settled the grammar of schooling by asking youth to explore collaboration outside of the school context, bringing youth participants to relatable, everyday spaces such as a 70 person student cooperative house. Inspired by these spaces, youth used junk material to construct ideal, fantastical collaborative spaces. Youth then considered how those ideal collaborative spaces could be made more engaging and joyful by technology-based tools. Finally, youth were given an opportunity to consider how those ideal collaborative spaces -- and the technologies within those space -- could be brought into the context of modern schools. 

We found that our Learning Futures Workshop supported youth in re-imagining key goals of schools and opened up expansive technological possibilities. Starting from the fantastical space, youth constructed a collaborative space where they could dream together about a future they could see themselves being in. Within this collaborative space, they imagined a technology that brought their dreams to life in real-time while they spoke. When bringing this ideal space into their actual classrooms, youth expressed concerns that their goal of dreaming would not be acceptable inside classroom space, where work is the primary goal. Encouraged by the facilitators, youth re-imagined a classroom space where the goals of school were both dreaming and work, supported by their dream-based technology. While the dreaming technology as imagined by the youth was not directly feasible given the existing state of the art, youth were encouraged that advancements in the state of the art in generative models could support such a tool in the near future. 

Our approach towards PAI holds promise for the \textit{research community} as well, where these expansive envisioning through lived experience of participants opens the door for ambitious horizons for the progress of AI.  

\section{Acknowledgements}

We are grateful to Thomas M. Philip, Arturo Cortez, and Ashieda McKoy, who were instrumental in the planning and facilitation of the Learning Futures Workshop. This research was made possible by NSF grant DRL-2019806, DRL-2019805, and NSF grant $\#$2021159. We would also like to thank the anonymous reviewers who provided very useful comments. Finally, the opinions expressed are those of the authors, and do not represent views of the
NSF.

\bibliographystyle{plainnat}
\bibliography{neurips.bib}

\end{document}